\documentclass[a4paper]{jpconf}
\usepackage{graphicx}
\usepackage{lineno}
\newcommand{\jpsi}{${\rm J/\psi}$~}
\newcommand{\pT}{$p_{ {\mathrm T} }~$}
\newcommand{\snn}{$\sqrt{s_{ {\mathrm {NN}} } }$~}
\newcommand{\ycms}{$y_{cms}$}

\begin{document}

\title{Inclusive \jpsi production in p-Pb collisions with ALICE at the LHC}

\author{Igor Lakomov for the ALICE Collaboration}

\address{Institut de Physique Nucl\'{e}aire d'Orsay, Universit\'{e} Paris-Sud, 91406 Orsay, France}

\ead{ilakomov@cern.ch}


\begin{abstract}
In 2013, inclusive \jpsi measurements in p-Pb collisions were performed at the LHC at \snn= 5.02 TeV, in order to measure the effects related to the cold nuclear matter on the \jpsi production. Their evaluation is important in order to be able to disentangle hot and cold nuclear matter effects in Pb-Pb collisions. In this contribution, results on the inclusive J/$\psi$ production in p-Pb collisions, in the dimuon decay channel at forward and backward rapidity, down to zero transverse momentum, are presented and compared to theoretical models.
\end{abstract}

\section{Introduction}
Production of charmonia, bound state of $c$ and $\bar{c}$ quarks, is an intense research activity, both experimentally and theoretically. The peculiar properties of some of the charmonium states, like their small size ($< 1$ fm) and strong binding energy (several hundred MeV), make them ideal probes of the strongly interacting deconfined matter, the so-called Quark-Gluon Plasma (QGP), formed in ultra-relativistic heavy-ion collisions. ALICE (A Large Ion Collider Experiment) \cite{ALICE} was designed to study heavy-ion collisions and the properties of the QGP. One of the signatures of the QGP formation is the so-called \jpsi ``melting" \cite{matsui} that results in a suppression of the \jpsi in heavy-ion collisions. The \jpsi suppression can be quantified in terms of the nuclear modification factor:
\begin{equation}\label{eq:RAA}
R_{\rm AA} = \frac{Y_{\rm AA}}{<T_{\rm AA}>\times\sigma_{\rm pp}^{\rm J/\psi}},
\end{equation}
where $Y_{\rm AA}$ is the invariant yield of the \jpsi measured in nucleus-nucleus (A-A) collisions, $T_{\rm AA}$ is the nuclear thickness function, $\sigma_{\rm pp}^{\rm J/\psi}$ is the \jpsi cross-section in pp collisions at the same energy.
If $R_{\rm AA}$ differs from unity nucleus-nucleus collisions cannot be considered as a simple superposition of elementary nucleon-nucleon collisions. ALICE measured $R_{\rm AA}^{0-80\%} = 0.545 \pm 0.032(stat.) \pm 0.083(syst.)$ in Pb-Pb collisions at \snn= 2.76 TeV \cite{alicepbpb}. The \jpsi suppression observed in this measurement was well described in terms of hot nuclear matter effects (related to QGP formation). However other effects, related to cold nuclear matter (CNM) are expected to play a role. They can be divided into three main groups:
\begin{itemize}
  \item {\bf Initial state effects.} Gluon shadowing \cite{shad} (or gluon saturation \cite{satur}) is an effect of depletion of low-momentum gluons in nuclei compared to protons. It is currently poorly constrained by data at the LHC energy.
  \item {\bf Nuclear absorption.} A \jpsi, or a pre-resonant $c\bar{c}$ pair can be dissociated by interacting with cold nuclear matter \cite{absorp}. At the LHC, due to the large Lorentz boost of the colliding nuclei at mid­‐ and forward rapidity in p-­Pb collisions the $c\bar{c}$ pair spends a very short time within cold nuclear matter. Therefore such effect (also known as nuclear absorption) is expected to be negligible.
  \item {\bf Coherent parton energy loss.} An effect related to the medium-induced radiative loss by the initial (partons) or final ($c\bar{c}$ pair) state \cite{arleo}.
\end{itemize}

To investigate the role of CNM effects and disentangle them from the hot nuclear matter effects in Pb-Pb collisions, p-Pb collisions were studied at the LHC at \snn = 5.02 TeV in 2013. First results are already submitted for publication \cite{aliceppb}.

\section{Analysis}
The ALICE detector allows to study the \jpsi production down to zero transverse momentum in two decay channels: dimuon at forward and dielectron at mid-rapidity. In this work, we will concentrate on the dimuon channel, studied in the muon spectrometer \cite{ALICE} with an acceptance $-4 < y_{lab} < -2.5$. Due to the LHC beam energy asymmetry, the nucleon-nucleon center-of-mass system is shifted by $\Delta y = 0.465$ in the direction of the proton beam with respect to the laboratory system. Two configurations were adopted, obtained by inverting the direction of the two beams in the machine. The measurements of the \jpsi production in ALICE were then performed in two different rapidity ranges: backward ($-4.46 < $\ycms$ < -2.96$) and forward ($2.03 < $\ycms$ < 3.53$) with integrated luminosities of 5.81 $\pm$ 0.19 nb$^{-1}$ and 5.03 $\pm$ 0.18 nb$^{-1}$, respectively.
This analysis is based on the dimuon trigger selection, requiring a coincidence of minimum bias (MB) interaction with two opposite sign muon tracks detected in the trigger chambers of the muon spectrometer. The MB trigger efficiency for non-single-diffractive events is $>99\%$ \cite{MBtrigger}. The dimuon trigger efficiency reaches $96\%$ for $p_{{\mathrm T},\mu}>$1.5 GeV/$c$. To reject single muons at the edge of the acceptance of the spectrometer, the condition of $-4<\eta_{\mu}<-2.5$ was required.

\begin{figure}[h]
\begin{minipage}{19pc}
\includegraphics[trim=1pc 0.87pc 3.5pc 32pc, clip=true, width=19pc]{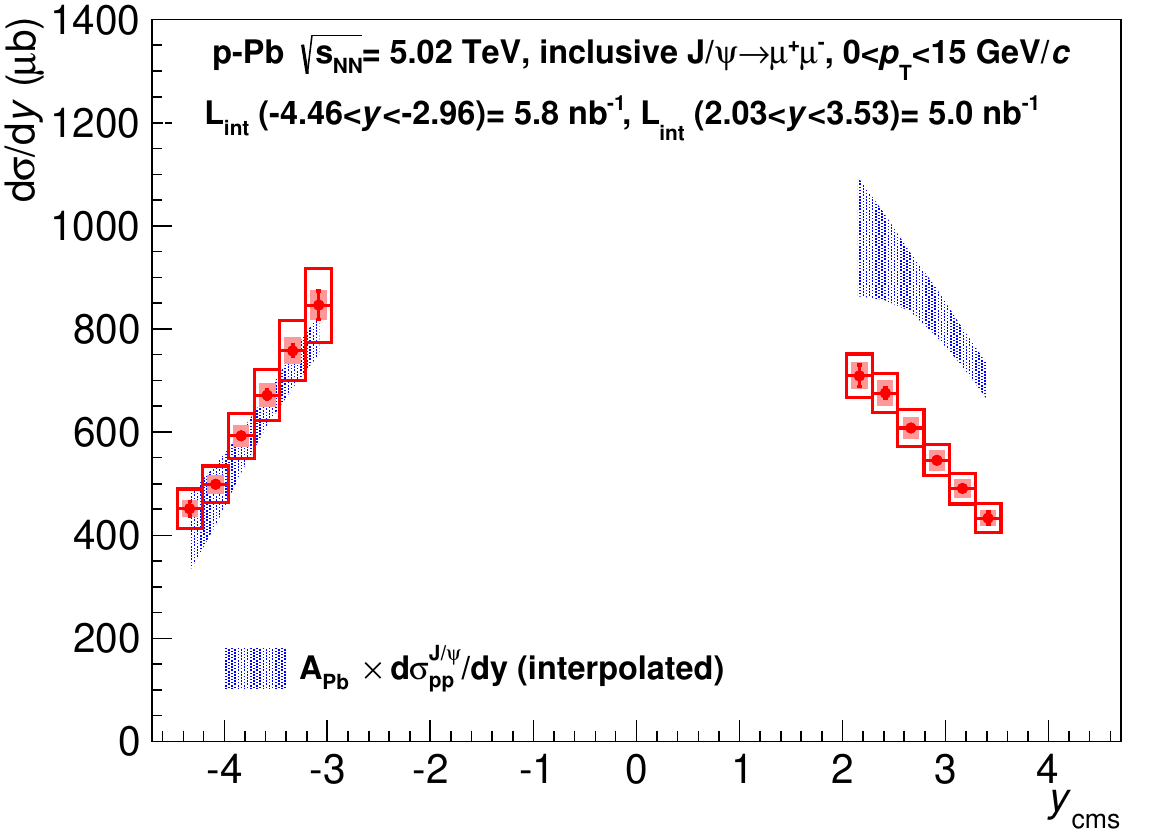}
\end{minipage}\hspace{0.1pc}
\begin{minipage}{19pc}
\includegraphics[trim=1pc 0.87pc 0pc 0.5pc, clip=true, width=19pc]{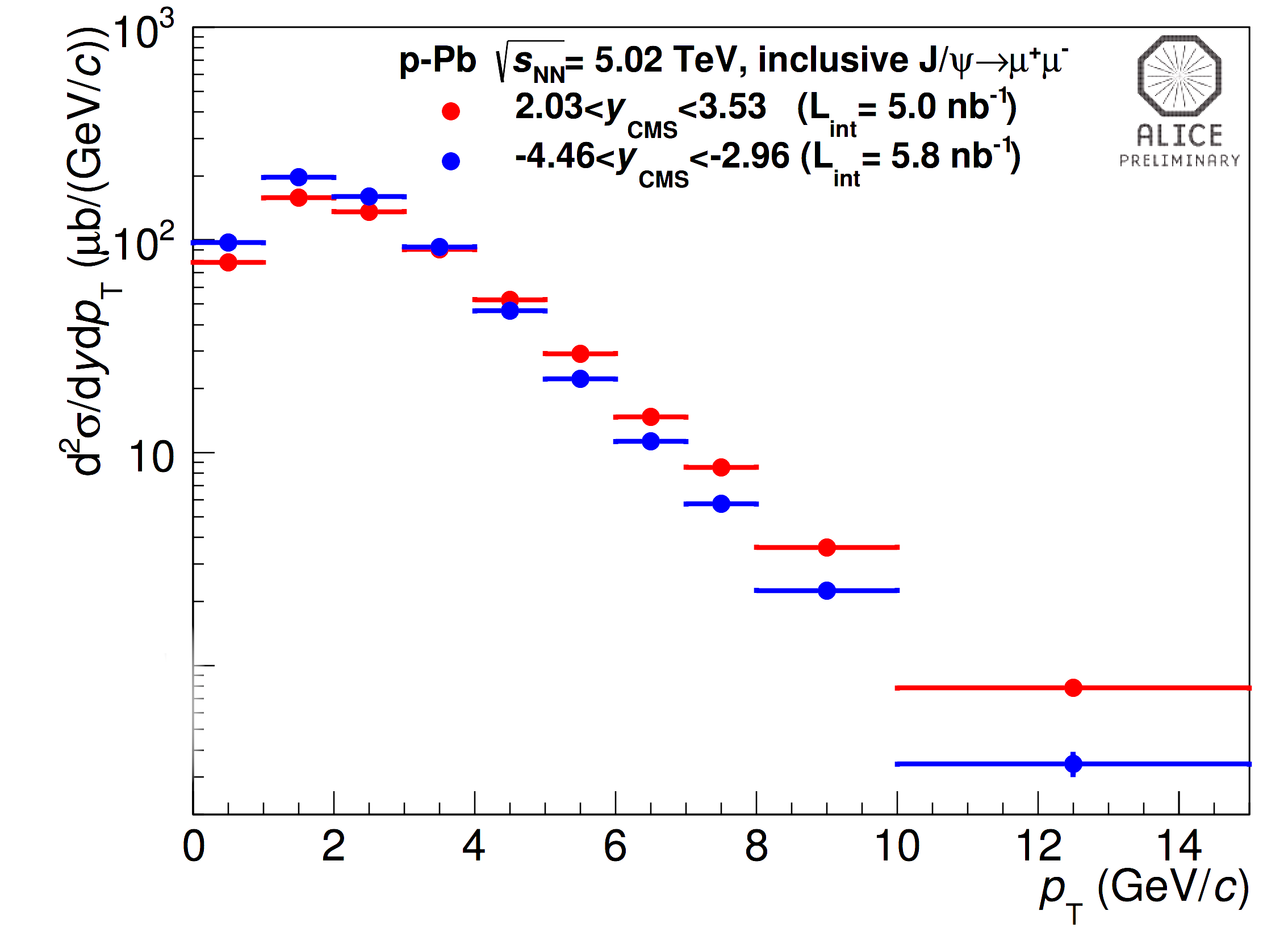}
\end{minipage}
\caption{\label{sigma}The inclusive \jpsi production cross section, as a function of rapidity and $p_{\mathrm T}$. Error bars represent statistical uncertainties, open boxes uncorrelated systematic uncertainties, shaded boxes show partially correlated systematic uncertainties. Bands in the left panel correspond to the result of the interpolation of the inclusive $\sigma_{\rm pp}^{\rm J/\psi}$ scaled by the Pb-nucleus mass number $A_{\rm Pb}$. Left panel is from \cite{aliceppb}.}
\end{figure}

The raw \jpsi yield was extracted by fitting the opposite-sign dimuon invariant mass spectrum by the sum of a background and a signal function. The numbers of \jpsi and its systematic uncertainty were evaluated from different fits varying background, signal (\cite{crystalball,NA60}) functions and the fitting range.
The acceptance times efficiency ($A\times \epsilon$) was estimated using a pure \jpsi signal Monte-Carlo simulation tuned on the real data kinematic distributions. The integrated values of $A\times \epsilon$ are ($17.1\pm 1.2$)\% and ($25.4\pm 1.3$)\% for the backward and forward rapidity regions, respectively. The difference in the values results from different tracking efficiency in the two periods of data taking.
In order to study the nuclear effects on the \jpsi production in p-Pb collisions, the nuclear modification factor $R_{\rm pPb}$ defined as (\ref{eq:RAA}) was used. Given the fact that there are no measurements of $\sigma_{\rm pp}^{\rm J/\psi}$ at $\sqrt{s} =$ 5.02 TeV, an interpolation procedure was followed, based on the following steps:
\begin{itemize}
  \item {\bf Energy interpolation:} existing ALICE pp results at $\sqrt{s}=2.76$ and 7 TeV (covering $2.5 < y < 4.0$ in six bins) were interpolated with empirical shapes (linear, power law and exponential).
  \item {\bf Rapidity interpolation/extrapolation:} due to the $y$-shift in p-Pb, the results from the previous step were fitted with empirical shapes (gaussian, polynomial) to reach the desired $y$-coverage. 
\end{itemize}
The results of the procedure were validated by comparison with theoretical models: CEM \cite{CEM} and actual FONLL calculation for $c\bar{c}$ production \cite{FONLL}. The values for $\sigma_{\rm pp}^{\rm J/\psi}$ scaled by the Pb-nucleus mass number are shown in the left panel of Fig.~\ref{sigma} by a blue band representing the uncertainties estimated by this procedure.

In addition to $R_{\rm pPb}$ the forward-to-backward ratio $R_{\rm FB}$, defined as:
\begin{equation}\label{eq:RFB}
R_{\rm FB} (|y_{cms}|) = \frac{R_{\rm pA}(y_{cms})}{R_{\rm pA}(-y_{cms})} = \frac{Y_{\rm pA}(y_{cms})}{Y_{\rm pA}(-y_{cms})}.
\end{equation}
and independent of the pp reference cross section, was also studied.

$R_{\rm FB}$ was measured in a narrower rapidity range ($2.96 < |$\ycms$| < 3.53$) than $R_{\rm pPb}$. The advantages of $R_{\rm FB}$ are that the nuclear thickness function and $\sigma_{\rm pp}^{\rm J/\psi}$ cancel out in the ratio and therefore, a pp interpolation is not needed. However the overall statistics were reduced by 30\%.

\begin{figure}[h]
\includegraphics[trim=1pc 0.87pc 4pc 32pc, clip=true, width=18pc]{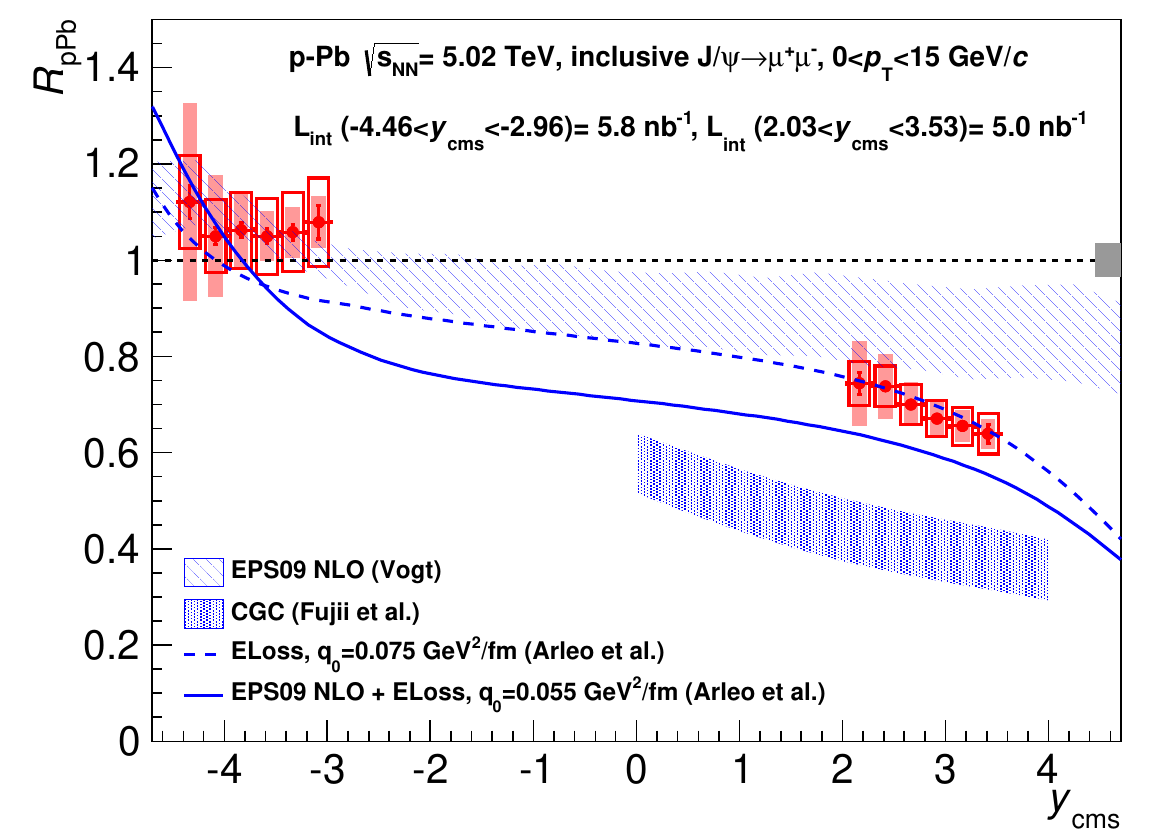}\hspace{3pc}
\begin{minipage}[b]{16pc}\caption{\label{RpA}The $y$-dependence of nuclear modification factor for inclusive \jpsi production at \snn = 5.02 TeV, compared to theorethical predictions \cite{vogt, arleo2, CGC}. Error bars correspond to statistical uncertainties, open boxes represent uncorrelated systematic uncertainties, shaded boxes show partially correlated systematic uncertainties. Box around $R_{\rm pPb} = 1$ shows correlated uncertainties. From \cite{aliceppb}.}
\end{minipage}
\end{figure}

\section{Results}
The measured \jpsi cross-sections in p-Pb collisions at \snn = 5.02 TeV for the full statistics are shown in Fig.~\ref{sigma} differentially in $y$ (left panel) and in $p_{\mathrm {T}}$ (right panel). In the backward region, the \jpsi production is larger, and softer in $p_{\mathrm {T}}$. The nuclear modification factors $R_{\rm pPb}$ integrated over \pT for the forward and backward rapidity ranges are:

$R_{\rm pPb}$ ($2.03<$\ycms$<3.53$) = $0.70 \pm 0.01^{stat.} \pm 0.04^{syst.uncorr.} \pm 0.03^{syst.part.corr.} \pm 0.03^{syst.corr.}$

$R_{\rm pPb}$ ($-­‐4.46<$\ycms$<-­‐2.96$)=$1.08 \pm 0.01^{stat.} \pm 0.08^{syst.uncorr.} \pm 0.07^{syst.part.corr.} \pm 0.04^{syst.corr.}$

The uncertainty is dominated by the pp interpolation and the tracking efficiency. In Fig.~\ref{RpA} the rapidity dependence of $R_{\rm pPb}$ is presented. In the forward rapidity region the Color Glass Condensate (CGC)-based model underestimates the data while predictions of models including parton energy loss and/or shadowing are in fair agreement with the data.

\begin{figure}[h]
\begin{minipage}{19pc}
\includegraphics[trim=1pc 0.87pc 3.2pc 32pc, clip=true, width=19pc]{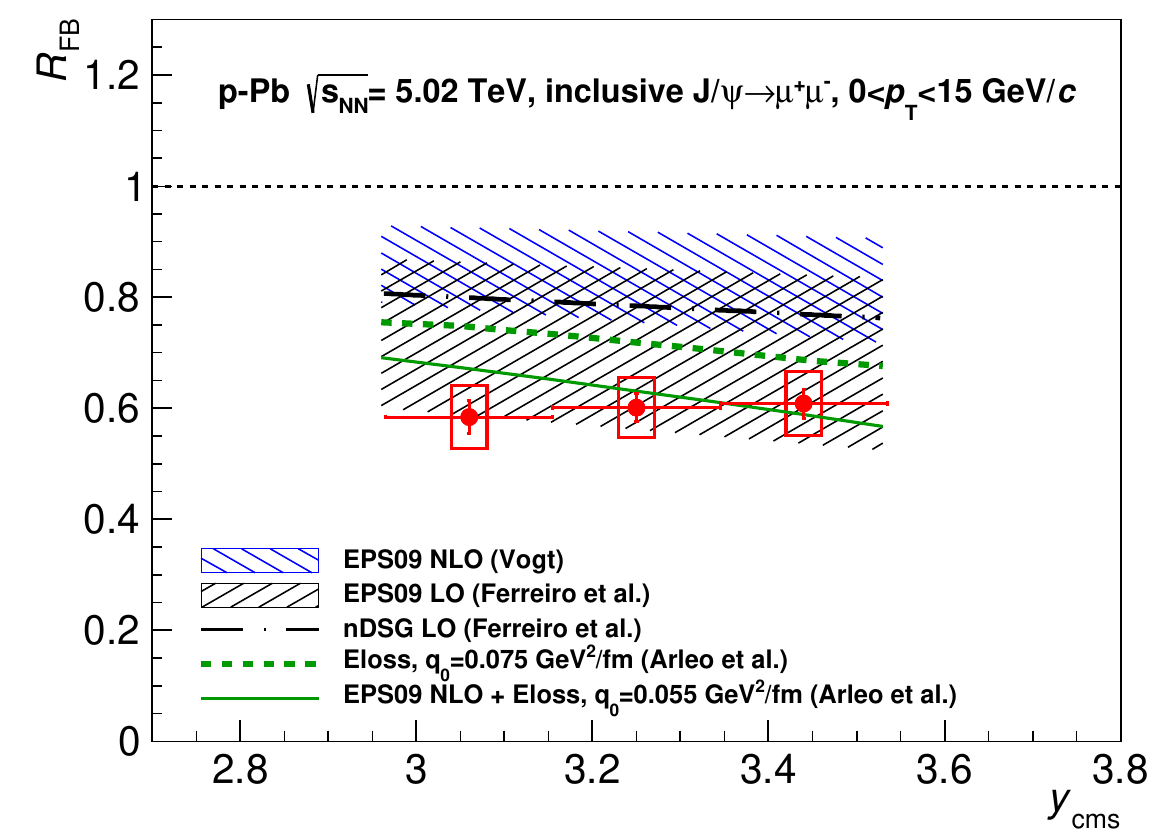}
\end{minipage}\hspace{0.1pc}%
\begin{minipage}{19pc}
\includegraphics[trim=1pc 0.87pc 3.5pc 31.5pc, clip=true, width=19pc]{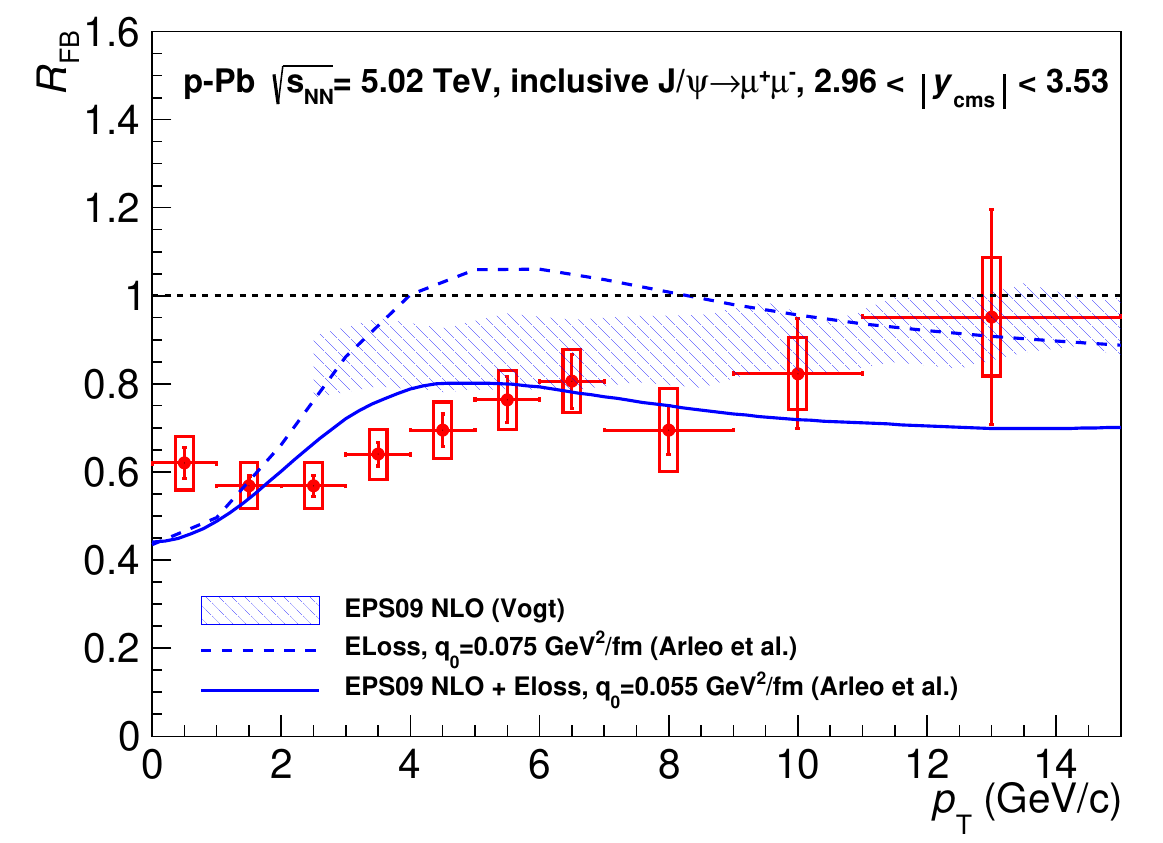}
\end{minipage}
\caption{\label{RFB}The forward to backward ratio $R_{\rm FB}$ for inclusive \jpsi production, as a
function of $y$ and $p_{\rm T}$, compared to theoretical models \cite{vogt, arleo2, CGC, ferreiro, arleo3}. Statistical uncertainties are shown as bars, uncorrelated systematic uncertainties correspond to open boxes. From \cite{aliceppb}.}
\end{figure}

An integrated value of $R_{\rm FB} = 0.60 \pm 0.01 (stat.) \pm 0.06 (syst.)$ was measured. In Fig.~\ref{RFB} the results on $R_{\rm FB}$ are shown as a function of rapidity (left) and $p_{\rm T}$ (right). The uncertainty on $R_{\rm FB}$ is dominated by the tracking efficiency. A model which includes shadowing as the only CNM effect slightly overestimates the data integrated over $p_{\mathrm {T}}$, while the inclusion of  an energy loss component leads to a better agreement. A sizeable $p_{\rm T}$ dependence of $R_{\rm FB}$ with a stronger suppression at low $p_{\rm T}$ is measured. Theoretical models including energy loss and shadowing show strong nuclear matter effects at low $p_{\rm T}$ in fair agreement with the data.

\section{Conclusions}
Inclusive \jpsi suppression was measured by ALICE in p-Pb collisions at \snn = 5.02 TeV. The $R_{\rm FB}$ ($R_{\rm pPb}$) were measured vs $y$ and \pT (vs $y$) and compared with theoretical models. Within uncertainties, both the shadowing and the coherent energy loss approaches give a fair description of the data, while the CGC-inspired model predicts too much suppression.

\end{document}